\documentstyle[12pt, psfig]{article}

\def\reference{\parskip 0pt\par\noindent\hangindent 0.5 truecm}

\begin{document}
%
%
\title{The Chemical Yields Produced by Zero Metal Massive Stars\footnote{Invited Talk, Annual
Conference of the Astronomical Society of Australia, Lorne, 2--4
July, 2001.}}
%


\author{Marco Limongi$^{1}$ and
 Alessandro Chieffi$^{1,2}$
} 

\date{}
\maketitle

{\center
$^1$ Osservatorio Astronomico di Roma, Via Frascati 33, Monteporzio Catone (Roma), Italy, I-00040\\marco@mporzio.astro.it\\[3mm]
$^2$ Istituto di Astrofisica Spaziale (CNR), Via Fosso del Cavaliere, Roma, Italy, I-00133\\achieffi@ias.rm.cnr.it\\[3mm]
}


\begin{abstract}

In this review we compare the three existing sets of theoretical
yields of zero metal massive stars available in the literature. We
also show how each of these three different sets of yields fits
the element abundance ratios observed in the extremely metal poor
star $\rm CD~38^{o}~245$. We find that, at present, no theoretical
set of yields of zero metal massive stars is able to
satisfactorily reproduce the elemental ratios [X/Fe] of this star.

\end{abstract}

{\bf Keywords:} stars: evolution --- nucleosynthesis ---
supernovae: general --- stars: abundances --- Galaxy: formation

\bigskip

%
%

\section{Introduction}
The astrophysical relevance of the first generation of massive
stars is certainly connected to the chemical enrichment of the
primordial interstellar medium. In fact, current Big Bang theories
predict that no metals were produced in a significant amount by
the Big Bang nucleosynthesis. If we couple this information with
the evidence that, at present, metals do exist in the universe and
with the current belief that metals are mainly synthesised in
stars, we cannot escape the conclusion that at a certain point in
the evolution of the universe, zero metal stars did form and that
the more massive ones were the first to enrich the pristine
material. In spite of this astrophysical relevance only few papers
discuss the evolution and nucleosynthesis of zero metallicity
massive stars. Woosley \& Weaver (1995, WW95 hereafter) presented
the yields of massive stars in the range $\rm 13-40~M_\odot$ for
five initial metallicities, namely, Z=0, 0.001, 0.01, 0.1, 1 $\rm
Z_\odot$, discussing the dependence of the yields on the initial
mass and metallicity. By the way let us remind the reader that the
yield of any given isotope is defined as the mass in solar masses
of that isotope ejected by the star. Chieffi, Limongi, Straniero,
\& Dominguez presented and discussed the evolutionary properties,
the explosions and the yields of zero metal stars in the range
$\rm 15-80~M_\odot$ in a series of papers in the last few years
(Limongi, Straniero \& Chieffi 2000; Chieffi et al.\ 2001a, b;
Limongi et al.\ 2001; Limongi \& Chieffi 2001). Finally Umeda \&
Nomoto (2002, UN02 hereafter) have recently published the yields
of zero metal stars in the range $\rm 13-30~M_\odot$.

The first aim of this paper is to compare these three sets of
computations and to underline differences and similarities among
them.

The second one is that of using these theoretical yields to fit
the surface chemical composition of extremely metal poor low mass
stars. Such a direct comparison is possible because these stars
probably formed in an environment enriched by just the first
generation of stars. Hence they give us a unique opportunity of
observing directly the ejecta of a single stellar generation and
not the complex superimposition of many generations of stars of
different metallicity (as happens when looking at stars of higher
metallicity). Moreover, recent sets of observational data
(McWilliam et al.\ 1995; Ryan, Norris, \& Beers 1996) have shown
that below $\rm [Fe/H]\simeq-2.5$ there exists a significant star
to star scatter in the observed element abundance ratios. This
scatter has been interpreted as a signature of the fact that these
stars formed in a highly inhomogeneous medium enriched by very few
supernovae (Auduze \& Silk 1995). In this scenario, each
primordial cloud was enriched by just one supernova (or, at most,
a mixture of two to three SN II) so that the low mass stars of the
second generation could preserve, up to the present time, the
chemical composition of matter enriched by a single zero metal
type II supernova. $\rm CD~38^{o}~245$ is one of the most metal
poor stars presently known and it is probably a good candidate for
being such a second generation star. In this paper we intend to
discuss mainly the method we intend to adopt to analyse these very
metal poor stars and hence we will discuss just the quoted star; a
complete analysis of the full sample of the very metal poor stars
presently known is in preparation and will be presented shortly.

\section{Comparison Between the Existing Sets of Yields Produced by Zero Metal Massive Stars}

The final chemical composition of the ejecta of a type II
supernova is the result of the combined effect of the
pre-supernova evolution and of the passage of the shock wave
through the mantle of the star during the explosion.

\begin{figure}[h]
\centerline{\psfig{file=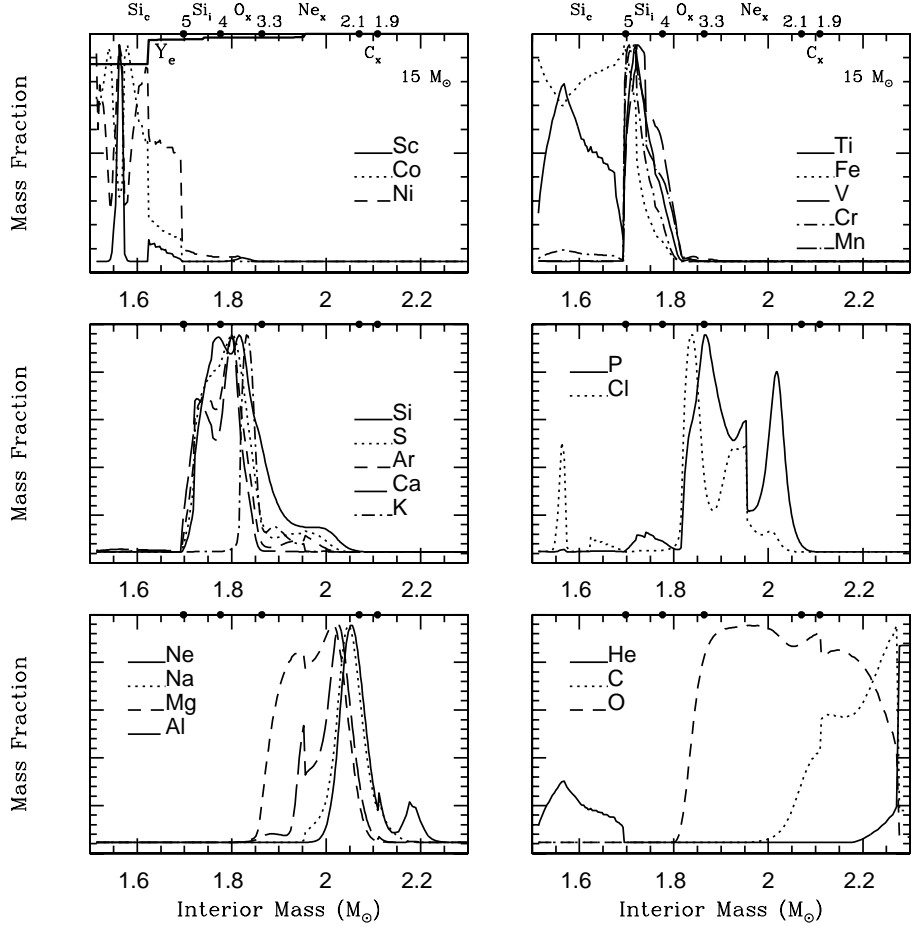,width=13cm}}
\caption{Chemical composition left by the passage of the shock wave in a $\rm 15~M_\odot$ model. See text for more details.}
\label{figlabel}            
\end{figure}

Figure 1 shows the chemical composition left by the passage of the
shock wave in a $\rm 15~M_\odot$ zero metallicity model taken as a
representative case. The x axis refers to the mass coordinate; the
black dots on the upper x axis in each panel mark the zones heated
up to a maximum temperature of 5, 4, 3.3, 2.1, and 1.9 billion
degrees during the explosion. These zones are the ones that
undergo complete explosive Si burning, incomplete explosive Si
burning, explosive O burning, explosive Ne burning, and explosive
C burning respectively. The zones more external in mass are left
untouched by the explosion. The various elements are plotted in
the same graph but using different limits for the y axis. Figure 1
clearly shows that each element is produced in a well defined zone
of the star after the explosion, so that they can be divided into
different groups depending on their production site. In particular
we can identify the following groups: (1) Sc ($\rm ^{45}Sc$, $\rm
^{45}Ca$), Co ($\rm ^{59}Co$), and Ni ($\rm ^{58}Ni$) are produced
by explosive complete Si burning; (2) Ti ($\rm ^{48}Ti$) and Fe
($\rm ^{56}Fe$) are produced by a combination of complete and
incomplete explosive Si burning; (3) Cr ($\rm ^{52}Fe$), V ($\rm
^{51}Cr$), and Mn ($\rm ^{55}Co$) are produced only by incomplete
explosive Si burning; (4) Si ($\rm ^{28}Si$), S ($\rm ^{32}S$), Ar
($\rm ^{36}Ar$), and Ca ($\rm ^{40}Ca$) are produced by a
combination of incomplete explosive Si burning and explosive O
burning; (5) K ($\rm ^{39}K$) is the only element produced
exclusively by explosive O burning; (6) Ne ($\rm ^{20}Ne$), Na
($\rm ^{23}Na$), Mg ($\rm ^{24}Mg$), Al ($\rm ^{27}Al$), P ($\rm
^{31}P$), and Cl ($\rm ^{35}Cl$, $\rm ^{37}Ar$) are produced in
the C convective shell during the hydrostatic evolution and then
partially modified during the explosion by explosive C/Ne burning;
(7) He ($\rm ^{4}He$), C ($\rm ^{12}C$), N ($\rm ^{14}N$), O ($\rm
^{16}O$), and F ($\rm ^{19}F$) are produced during the hydrostatic
evolution and left untouched by the explosion. Elements pertaining
to the first two groups depend significantly on the location of
the mass cut, i.e.\ the mass coordinate which separates the
remnant from the ejecta, because they are produced in the more
internal zones. By the way, the mass cut is still a highly
uncertain quantity due to the lack of self consistent models
leading to a successful explosion for type II supernovae. All the
other elements are not affected by the exact location of the mass
cut.

\begin{figure}[h]
\centerline{\psfig{file=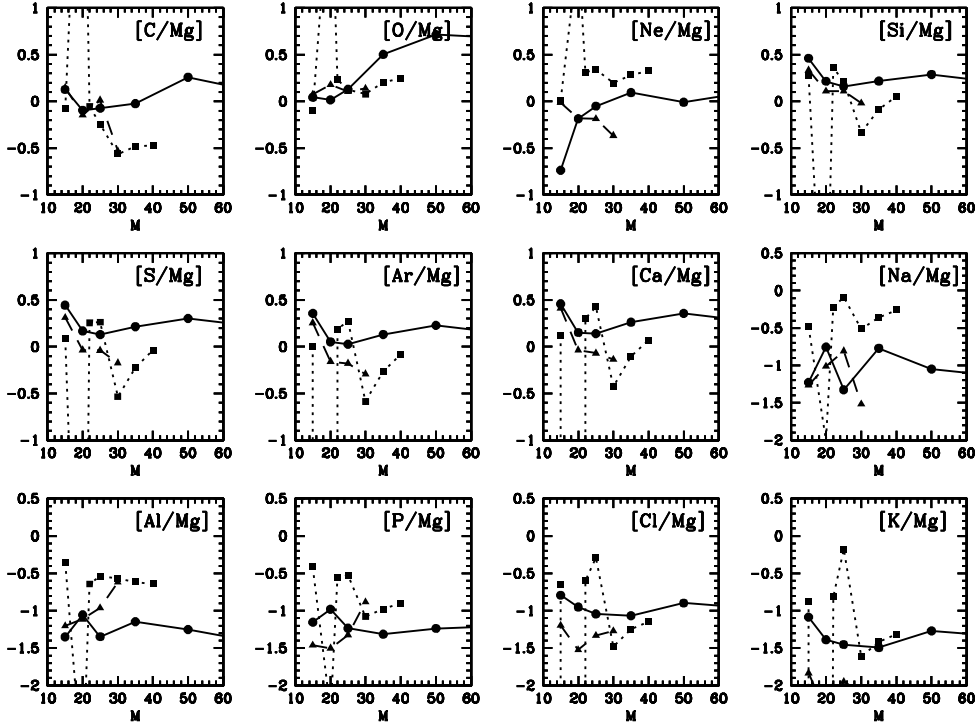,height=10cm}}
\caption{Trend with the mass of the [X/Mg] log ratios obtained by
using the three sets of theoretical yields: LC01 ({\em filled
circles} connected by the {\em solid line}); WW95 ({\em filled
squares} connected by the {\em dotted line}); UN02 ({\em filled
triangles} connected by the {\em dashed line}). }
\label{figlabel}            
\end{figure}

Figure 2 shows the trend with the mass of the [X/Mg] ($\rm \equiv
Log(X/Mg)-Log(X/Mg)_\odot$) log ratios for the elements that are
only marginally influenced by the location of the mass cut. The
elements are ordered on the basis of their production site, i.e.\
the first three panels in the first row refer to the elements
produced by hydrostatic burning (i.e.\ C, O, and Ne); the next
four panels refer to elements produced by both incomplete
explosive Si and explosive O burning (i.e.\ Si, S, Ar, and Ca);
the following four panels refer to elements produced during the
hydrostatic evolution of the stars in the C convective shell and
then partially modified by the explosion (i.e.\ Na, Al, P, and
Cl); the last panel refers to the single element produced by
explosive O burning, i.e.\ K. The filled circles connected by a
solid line refer, in each panel, to our computations (see Tables
1--6, LC01 hereafter), the filled squares connected by a dotted
line refer to the yields computed by WW95 (their case C if
present), and the filled triangles connected by a dashed line to
the yields by UN02. Note that the yields produced by the
$20~M_\odot$ computed by WW95 are largely altered by the presence
of a very extended fall back: this occurrence explains the
presence of a `spike' in this figure and in the following ones.
Instead of comparing the yields (in solar masses) ejected by any
stellar model, we preferred to compare a quantity that can be
directly observed in real stars. We chose the [X/Mg] log ratios
because Mg does not depend on the mass cut; it has a rather strong
dependence on the initial mass and  is also well determined
observationally. All the panels in Figure 2 show that the yields
computed by the three groups differ significantly. Since the final
abundances of these elements depend mainly on the pre-supernova
evolution, it is clear that there is some basic difference in the
computation of the evolution of these stars (e.g.\ treatment of
convection, $\rm ^{12}C(\alpha,\gamma)^{16}O$ rate and so on).
Unfortunately it is not easy to understand where these differences
come from because neither WW95 nor UN02 published any detail (up
to now) about their pre-supernova models. There are however a few
general comments about the yields, and their trends, worth
mentioning. First of all, both our and UN02 yields show a more or
less continuous dependence on the mass while the WW95 yields show
a rather scattered behaviour in which it is not easy to identify a
clear dependence on the mass. Second, both WW95 and UN02 predict
[O/Mg] of the order of zero over all the mass intervals they
explore, while we predict a trend of this ratio with the initial
mass which reaches a maximum value of the order of 0.7 dex; this
well defined difference could constitute a good observational
check for the models. A further consideration worth noting is the
behaviour of the four $\alpha$ elements Si, S, Ar, and Ca: though
the three groups predict different yields, within each set of
computations all four elements show exactly the same dependence on
the initial mass. This is a consequence of the fact that these
four elements are strongly coupled to each other because they are
produced by the same explosive burning: hence their internal
ratios (e.g. [Si/S], [S/Ca], [Ar/Si] and the like) are largely
independent of the mass, the initial chemical composition, and the
author. By the way, let us remind the reader also that the
thermonuclear supernovae produce these four elements in the same
relative proportions: this is simply due to the fact that the
physical conditions in which these four elements are produced are
very similar in the thermonuclear and core collapse supernovae
(i.e.\ the incomplete explosive Si burning and explosive O burning
in an environment only marginally neutronised). Let us eventually
note that WW95 predict a much larger ratio for the odd elements,
Na, Al, P, than we and UN02 do.

\begin{figure}[h]
\centerline{\psfig{file=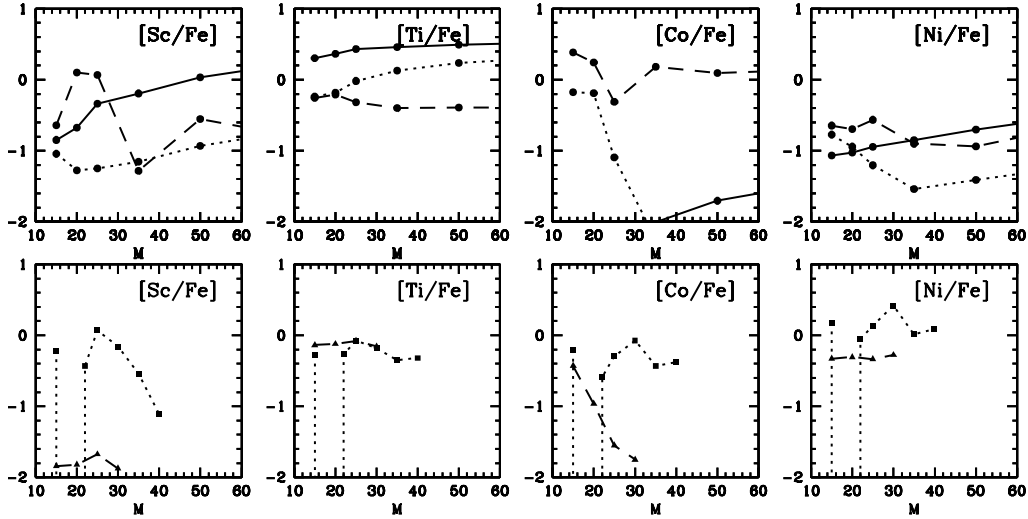,height=8cm}}
\caption{Trend with the mass of [Sc,Ti,Co,Ni/Fe] log ratios for
the three sets of yields. The LC01 yields ({\em filled circles})
are plotted in the {\em upper panels} for three different values
of the $\rm ^{56}Ni$ ejected, i.e.\ $\rm 0.01~M_\odot$ ({\em solid
line}), $\rm 0.1~M_\odot$ ({\em dotted line}) and the one obtained
when the mass cut coincides with the iron core ({\em dashed
line}). The WW95 and UN02 yields are plotted in the lower panels;
the {\em filled squares} connected by the {\em dotted line} refer
to the WW95 yields while the {\em filled triangles} connected by
the {\em dashed line} refer to the UN02 yields.}
\label{figlabel}            
\end{figure}

Figure 3 shows the trend with the initial mass of the elements
which are produced by the complete and incomplete explosive Si
burning. In this case the adopted reference element is Fe. These
elements are produced in the deep interior of the star hence their
yields will largely depend on the mass cut. Since this is a very
uncertain parameter, we decided to compute yields for various
values of the mass cut. The solid, dotted, and short dashed lines
in the first row of Figure 3 show our predictions for three
different choices of the mass cut: the solid and dotted lines are
obtained by assuming that all stars eject an amount of $\rm
^{56}Ni$ equal to, respectively, $0.01~M_\odot$ and $0.1~M_\odot$.
The dashed line, on the other hand, has been obtained by imposing
the mass cut to coincide with the Fe core mass for each star in
the sample. Since both the generic element (Sc, Ti, Co, or Ni) and
the Fe depend on the mass cut, even a very large change in the
amount of $\rm ^{56}Ni$ ejected does not lead to extreme changes
in these ratios. [Sc/Fe] shows a behaviour which depends on the
mass cut but remains in any case within the range $\rm -1.3\leq [Sc/Fe] \leq 0.1$
dex. Also the range of values for [Ti/Fe] remains confined within
1 dex, i.e. between 0.5 and --0.4 for any choice of the mass cut.
[Ni/Fe] shows a similar behaviour as well, and it ranges between
--0.6 and --1.6 at most. [Co/Fe] is the only exception since it
shows a very strong dependence on the mass cut. The solid and
dotted lines in the second row of Figure 3 represent,
respectively, the WW95 and UN02 predictions. Both these groups did
not provide yields for different choices of the mass cut so that
it is not possible to show the dependence of their result on the
mass cut. The [Sc,Ti,Co/Fe] predicted by WW95 are compatible with
our values in the sense that, at least, they fall within the range
of possible values we predict, while their predictions for [Ni/Fe]
fall outside the range of values we obtain. UN02 predict [Sc/Fe]
and [Co/Fe] very low but  [Ti/Fe] and  [Ni/Fe] in close agreement
with WW95.

\begin{figure}[h]
\centerline{\psfig{file=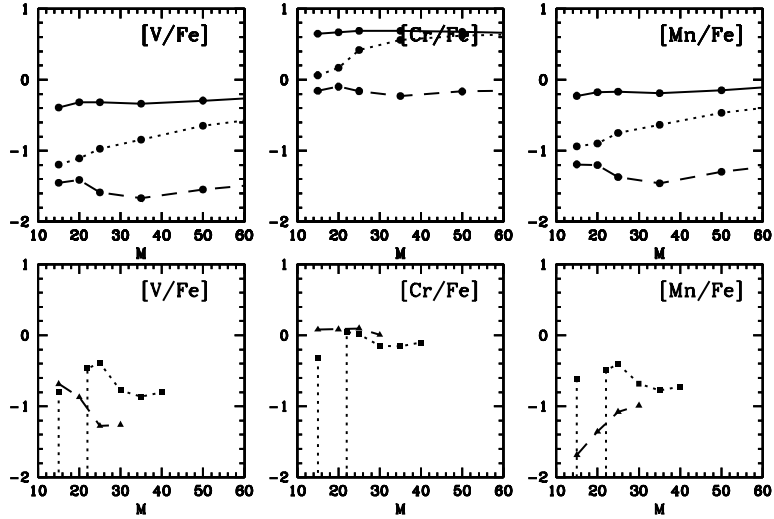,height=8cm}}
\caption{Trend with the mass of [V,Cr,Mn/Fe] log ratios for the
three sets of yields. The LC01 yields ({\em filled circles}) are
plotted in the {\em upper panels} for three different values of
the $\rm ^{56}Ni$ ejected, i.e.\ $\rm 0.01~M_\odot$ ({\em solid
line}), $\rm 0.1~M_\odot$ ({\em dotted line}), and the one
obtained when the mass cut coincides with the iron core ({\em
dashed line}). The WW95 and UN02 yields are plotted in the lower
panels; the {\em filled squares} connected by the {\em dotted
line} refer to the WW95 yields while the {\em filled triangles}
connected by the {\em dashed line} refer to the UN02 yields.}
\label{figlabel}            
\end{figure}

Figure 4 shows the behaviour of the three elements which are
produced by the incomplete explosive Si burning. Also in this case
the first row refers to the yields obtained for the three
different choices of the mass cut. The solid and dotted lines
refer to the case in which all stars eject an amount of $\rm
^{56}Ni$ equal to, respectively, $0.01~M_\odot$ and $0.1~M_\odot$,
while the dashed line refers to the case in which the mass cut
coincides with the Fe core mass. Though the three ratios depend on
the adopted mass cut, we can identify, for each panel, a permitted
and a forbidden region. A comparison with the WW95 and UN02
predictions (shown in the second row of Figure 4) shows that there
is a generic `compatibility' in the sense that their data fall
within our permitted region.

\section{Comparison Between Theoretical Yields and Observational Data}

The usual technique adopted to compare the observed and the
predicted [X/Fe] consists in shifting vertically the theoretical
predictions until a best fit to the data is obtained. This
procedure corresponds (in practice) to an artificial changing of
the amount of Fe ejected by the star after the explosion. Such a
procedure is essentially correct if it is applied to elements
which do not depend on the location of the mass cut while it may
be completely wrong if applied to elements whose yields depend on
it. The reason is that, in general, if we take the ejecta of a
given massive star, the relative scaling of all the various
elements is not necessarily preserved by changing the location of
the mass cut. In particular, the elements may be divided in two
groups: the first one (formed by elements from C to Ca) includes
all the elements which (presumably) are not largely affected by
the uncertainty in the mass cut location. Hence, a changing of the
mass cut will simply shift simultaneously the [X/Fe] of these
elements by the same amount without modifying their relative
abundances. The second one (formed by elements from Sc to Ni)
includes elements which strongly depend on the location of the
mass cut. In this case changing this parameter will not preserve
the relative scaling of the [X/Fe] of these elements since both
the Xs and the Fe depend (not necessarily in the same way) on it.
It is therefore clear that a `proper' choice of the mass cut is
crucial in this comparison. Since we are assuming that these metal
poor stars formed in an environment enriched by just one
supernova, once the exploding mass has been fixed, we can (almost
always) directly derive the amount of Fe ejected by requiring the
fit to a given observed [X/Fe] log ratio. We chose to fit the
observed [Mg/Fe]. Of course, the amount of Fe determined in this
way will depend on the mass of the exploding star because the
amount of Mg itself is a function of the initial mass.

Just as an example of the method we will present, in the
following, the comparison between the three existing sets of
yields and the element abundance pattern observed in the star $\rm
CD~38^{o}~245$ ([Fe/H]=-4.01) which is the most metal poor star in
the McWilliam et al.\ (1995) database.

\begin{figure}[h]
\centerline{\psfig{file=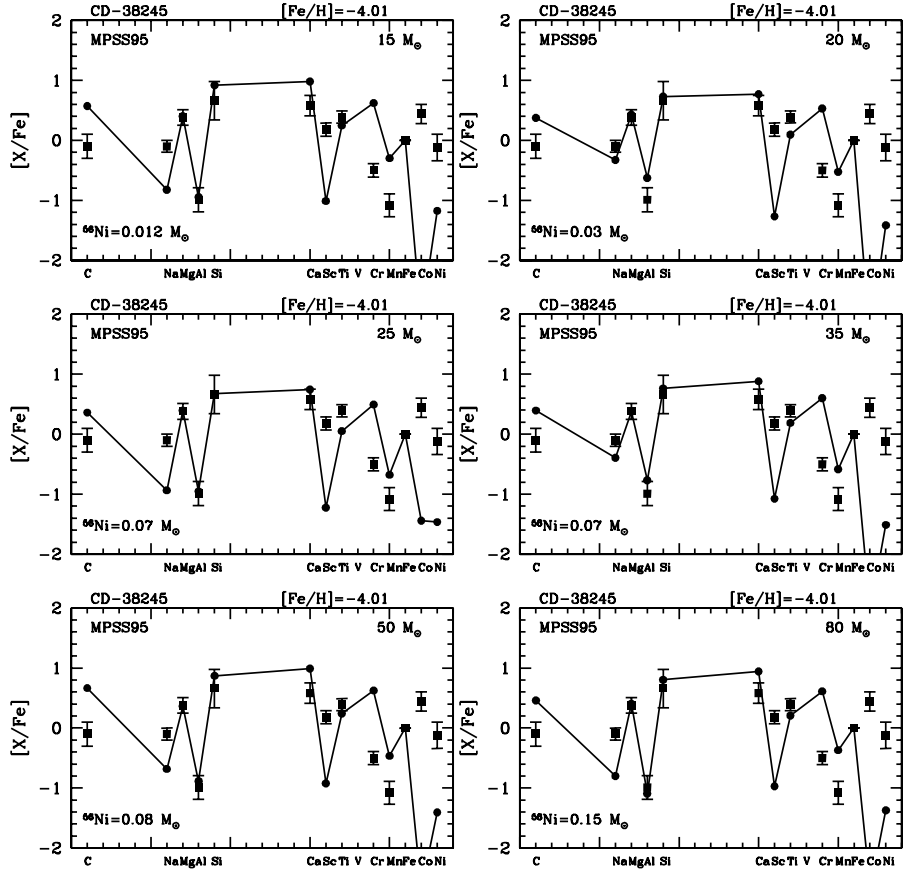,height=13cm}}
\caption{Fit to the $\rm CD~38^{o}~245$ with LC01 yields. See text for more details.}
\label{figlabel}            
\end{figure}

Figure 5 shows the fit to $\rm CD~38^{o}~245$ with our theoretical
yields. The filled squares represent the observed abundances while
the filled dots refer to the predicted abundances. The six panels
show the comparison between the quoted star and the ejecta of
different stellar masses (once the mass cut for each stellar mass
has been chosen to fit the observed [Mg/Fe]). Since the amount of
Mg increases significantly with the mass of the star, the amount
of $\rm ^{56}Ni$ ejected increases with the initial mass. The six
panels show that, once [Mg/Fe] is fitted, [Al/Fe] and [Si/Fe] are
also rather well reproduced by all the stellar models (which
actually span a wide mass interval). This is a consequence of the
fact that [Al/Mg] and [Si/Mg] depend very weakly on the initial
mass. This means that we cannot use these two ratios to
discriminate the mass of the exploding star but we can only say
that our pre-supernova models do reproduce them. Also [Ca/Fe] is
always within two sigma from the observed value, though the 20 and
$\rm 25~M_\odot$ models are those which provide the best fit to
the data. [C/Fe] and [Na/Fe] are never satisfactory reproduced,
the first of the two being always overestimated and the second one
being systematically underestimated: the 20 and the $\rm
35~M_\odot$ models are those for which the discrepancies are
minimised. A possible way to reconcile the fit of both  [C/Fe] and
[Na/Fe] with the data could be by slightly reducing the amount of
C left by the central He burning (e.g.\ by increasing the rate of
the $\rm ^{12}C(\alpha,\gamma)^{16}O$ reaction) since the effect
of such a change goes in the right direction (see Imbriani et al.\
2001). An extended set of computations would anyway be required to
support such a possibility. Looking at the iron peak elements,
only one of this group, i.e.\  Ti, is relatively well predicted by
the models (the 35 and $\rm 50~M_\odot$ models in particular): all
the other elements are missed by a substantial amount by all the
models in our database. Note that the discrepancies between the
theoretical predictions and the data are essentially independent
of the mass of the exploding star: Sc, Co, and Ni are always
systematically underestimated while Cr and Mn are always
overestimated.

\begin{figure}[h]
\centerline{\psfig{file=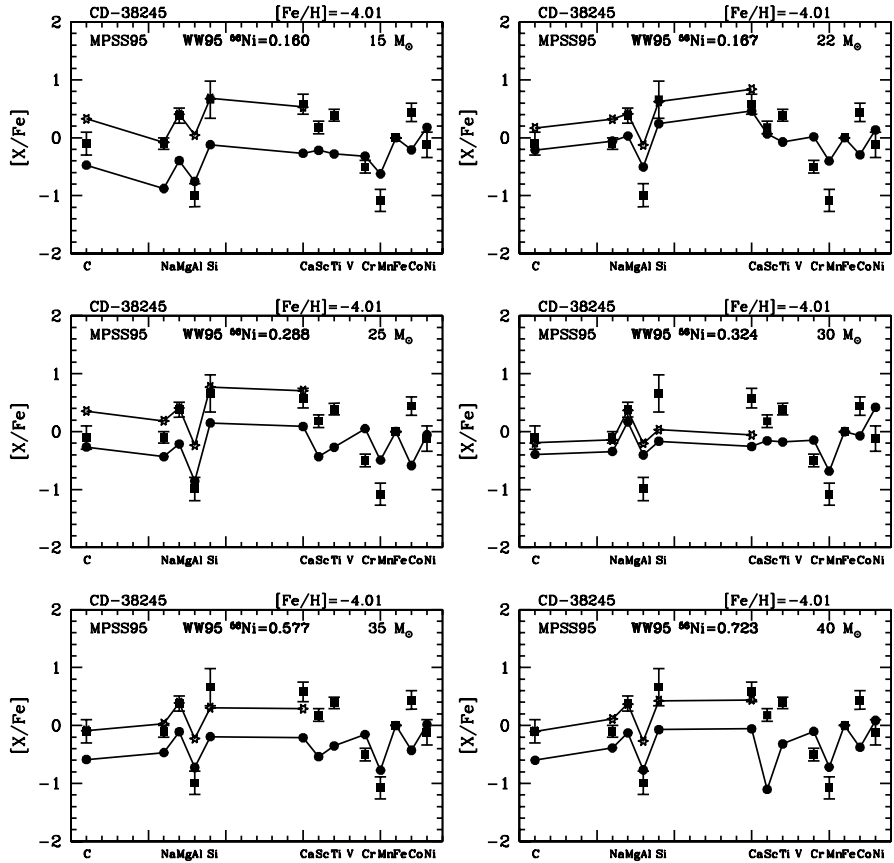,height=13cm}}
\caption{Fit to the $\rm CD~38^{o}~245$ with WW95 yields. See text for more details.}
\label{figlabel}            
\end{figure}

Figure 6 is similar to Figure 5 but for the WW95 yields. Since
WW95 give their yields for only one value of the $\rm ^{56}Ni$
ejected we cannot arbitrarily change the mass cut in order to fit
the observed [Mg/Fe]. However, since the abundance pattern of the
elements from C to Ca does not significantly depend on the
location of the mass cut, we can arbitrarily vertically shift all
the theoretical predictions of these elements (open stars in
Figure 6) to fit the observed [Mg/Fe] log ratio. By doing that we
find that [Al/Fe] is always overestimated by the models while
[Si/Fe] is always compatible with the observations except in the
30 and $\rm 35~M_\odot$ models. This means that, at variance with
our results, the relative scaling of Mg, Al, and Si is not
independent of the initial mass, even if it does not show a
continuous trend with the progenitor mass. [Ca/Fe] behaves like
[Si/Fe]. The only cases in which both [C/Fe] and [Na/Fe] are
simultaneously in agreement with the observations are  30 and
 $\rm 35~M_\odot$  which are, unfortunately, the only masses
that do not reproduce the observed [Si/Fe] and [Ca/Fe]. We cannot
make any comment on the comparison between the data and the
theoretical yields of the iron peak elements because we cannot
predict how their relative scaling changes by choosing the amount
of $\rm ^{56}Ni$ ejected needed to fit the observed [Mg/Fe].

\begin{figure}[h]
\centerline{\psfig{file=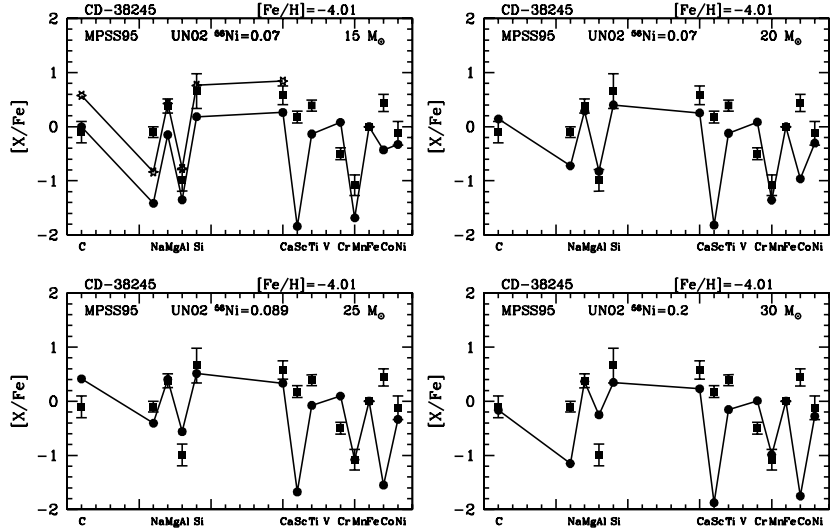,height=8cm}}
\caption{Fit to the $\rm CD~38^{o}~245$ with UN02 yields. See text for more details.}
\label{figlabel}            
\end{figure}

Figure 7 is similar to Figures 5 and 6 but for the UN02 yields.
Also UN02 give their yields for just one choice of the mass cut.
Also in this case we can vertically shift the abundances of all
the elements from C to Ca to fit the observed [Mg/Fe], if
necessary. The only model for which we are forced to shift the
theoretical yields is the $\rm 15~M_\odot$, in all the other cases
the amount of $\rm ^{56}Ni$ ejected is already the one able to fit
the observed [Mg/Fe]. Similar to the WW95 case, the relative
scaling of Mg, Al, and Si changes with the progenitor mass; in
this case, however, there seems to be a continuous trend with the
initial mass, i.e.\ [Al/Fe] increases while [Si/Fe] remains
roughly constant with increasing  progenitor mass (except the $\rm
15~M_\odot$ case). The final net result is that the only models in
which [Mg,Al,Si/Fe] agree simultaneously with the observations are
the 15 and the $\rm 20~M_\odot$. On the contrary the only model
which reproduces the observed [Ca/Fe] is the $\rm 25~M_\odot$.
There is no model able to simultaneously fit the observed [C/Fe]
and [Na/Fe]; in particular [Na/Fe] is always underestimated by all
the models, while [C/Fe] is fitted only by the 20 and $\rm
30~M_\odot$ models. Looking at the iron peak elements, the only
element abundance ratios reproduced by the UN02 models are [Mn/Fe]
and [Ni/Fe], all the other ones being missed by a significant
amount. In particular [Sc,Ti,Co/Fe] are systematically
underestimated while [Cr/Fe] is always overestimated.

\section{Conclusions}
We have compared the three existing sets of theoretical yields
produced by zero metallicity massive stars, i.e.\ the ones by
LC01, WW95, and UN02. The results may be summarised as follows: 1)
in general the yields computed by the three groups differ
significantly; 2) the [X/Mg] (X=C, O, Ne, Na, Al, Si, P, S, Cl,
Ar, K, Ca) obtained by LC01 and UN02 show a more or less
continuous dependence on the progenitor mass while the ones
obtained by WW95 show a much more scattered behaviour (no clear
dependence on the initial mass); 3) both WW95 and UN02 predict an
$\rm [O/Mg]$ almost flat around 0.0 dex in the mass range $\rm
13-40~M_\odot$; however we obtain a value which increases from 0.0
dex to 0.7 dex in the mass range $\rm 15-50~M_\odot$: this could
be a good observational check for the models; 4) within each set
of yields, [Si,S,Ar,Ca/Mg] show the same dependence on the mass,
i.e.\ their internal ratios (e.g.\ [Si/S], [S/Ca], [Ar/Si], and
the like) are largely independent of the initial mass; 5) WW95
predict a much lower odd--even effect (especially for Na, Al, and
P) than LC01 and UN02; the reason for this is difficult to
understand at present; 6) the abundance pattern of most of the
iron peak elements obtained by WW95, i.e.\ [V,Cr,Mn,Sc,Ti,Co/Fe],
is compatible with our results for a proper choice of the amount
of $\rm ^{56}Ni$ ejected, while [Ni/Fe] is always larger than the
range of values we obtain; 7) the abundance ratios of most of the
iron peak elements obtained by UN02, i.e.\ [V,Cr,Ti,Co/Fe],  are
compatible with our results for a proper choice of the mass cut,
while [Sc/Fe] and [Ni/Fe] always show large differences with
respect to our results.

A comparison between the three sets of yields and the element
abundance ratios observed in $\rm CD~38^{o}~245$ shows that the
observations are never satisfactorily reproduced by the models. In
particular, using the LC01 yields we find that: 1) the relative
scaling of C and Na are never reproduced by the models; 2)
[Mg,Al,Si,Ca/Fe] are always compatible with the observations
regardless of the progenitor mass (except in the $\rm 50~M_\odot$
case in which [Ca/Fe] is overestimated by the model); 3) the iron
peak elements are always missed by the models except [Ti/Fe] which
is always in very good agreement with the observations. Using the
WW95 yields we find that: 1) [Mg,Si,Ca/Fe] are always compatible
with the observations except for 30 and $\rm 35~M_\odot$ which
are, however, the only models which fit both [C/Fe] and [Na/Fe];
2) [Al/Fe] is always overestimated by the models; 3) no comment is
possible for the iron peak elements because we don't know how
their relative scaling changes by varying the $\rm ^{56}Ni$ to the
value needed to fit the observed [Mg/Fe]. Using the UN02 yields we
find that: 1) the abundance pattern of [Mg,Al,Si,Ca/Fe] is
compatible with the observations for the lower mass models, i.e.\
 15 and $\rm 20~M_\odot$; 2) [C,Na/Fe] are never simultaneously
in agreement with the observations; 3) the iron peak elements are
missed by all the models except [Mn/Fe] and [Ni/Fe] that are
better reproduced by the more massive stars, i.e.\  25 and $\rm
30~M_\odot$.

The final conclusion of this preliminary analysis is that at
present no theoretical set of yields can satisfactory reproduce
the surface chemical composition of $\rm CD~38^{o}~245$ under the
assumption that it is formed by material enriched by the ejecta of
just one zero metallicity type II supernova.



%
%





\section*{Acknowledgments}
One of us (ML) thanks the scientific committee of the 2001
Conference of the Astronomical Society of Australia for having
invited him to the meeting, and John Lattanzio and Brad Gibson for
their very generous hospitality during his visit in Australia.


\section*{References}






\reference Auduze, J., \& Silk, J.\ 1995, ApJ, 451, L49

\reference Chieffi, A., Limongi, M., Dominguez, I., \& Straniero,
O.\ 2001a, in New Quests in Stellar Astrophysics: The Link between
Stars and Cosmology (Dordrecht: Kluwer), in press

\reference Chieffi, A., Limongi, M., Dominguez, I., \& Straniero,
O.\ 2001b, in Chemical Enrichment of Intracluster and
Intergalactic Medium, ASP Conference Series, ed.\ F.\ Matteucci,
\& F.\ Giovannelli (San Francisco: ASP), in press

\reference Limongi, M., Straniero, O., \& Chieffi, A.\ 2000, ApJS,
129, 625

\reference Limongi, M., Chieffi, A., Straniero, O., \& Dominguez,
I.\ 2001, in New Quests in Stellar Astrophysics: The Link between
Stars and Cosmology (Dordrecht: Kluwer), in press

\reference Limongi, M., \& Chieffi, A.\ 2001, in Chemical
Enrichment of Intracluster and Intergalactic Medium, ASP
Conference Series, ed.\ F.\ Matteucci, \& F.\ Giovannelli (San
Francisco: ASP), in press

\reference McWilliam, A., Preston, G.W., Sneden, C., \& Searle,
L.\ 1995, AJ, 109, 2757

\reference Ryan, S.G., Norris, J.E., \& Beers T.C.\ 1996, ApJ,
471, 254

\reference Umeda, H., \& Nomoto, K.\ 2002, ApJ, 565, 385 (UN02)

\reference Woosley, S.E., \& Weaver, T.A.\ 1995, ApJS, 101, 181
(WW95)


\begin{table*}
\caption{\bf Yields of the $\rm \bf 15~M_\odot$ Z=0 model.}
\begin{tabular}{lccccccc}
\hline
 $\rm M(^{56}Ni)_{}^{}$  &    0.0010  &    0.0050  &    0.0100  &    0.0500  &    0.0750  &    0.1000 &     0.1160      \\      
\hline
 $\rm M_{\rm cut}$  &      1.81  &      1.78  &      1.77  &      1.71  &      1.68  &      1.63 &       1.60      \\
 $\rm M_{\rm ejected}$   &     13.19  &     13.22  &     13.23  &     13.29  &     13.32  &     13.37 &      13.40      \\
 H     &  7.73E+00  &  7.73E+00  &  7.73E+00  &  7.73E+00  &  7.73E+00  &  7.73E+00 &   7.73E+00      \\
 He    &  4.66E+00  &  4.66E+00  &  4.66E+00  &  4.66E+00  &  4.66E+00  &  4.66E+00 &   4.66E+00      \\
 C     &  1.40E-01  &  1.40E-01  &  1.40E-01  &  1.40E-01  &  1.40E-01  &  1.40E-01 &   1.40E-01      \\
 N     &  5.62E-07  &  5.62E-07  &  5.62E-07  &  5.62E-07  &  5.62E-07  &  5.62E-07 &   5.62E-07      \\
 O     &  3.16E-01  &  3.16E-01  &  3.16E-01  &  3.16E-01  &  3.16E-01  &  3.16E-01 &   3.16E-01      \\
 F     &  4.60E-11  &  4.60E-11  &  4.60E-11  &  4.60E-11  &  4.60E-11  &  4.60E-11 &   4.60E-11      \\
 Ne    &  7.87E-03  &  7.87E-03  &  7.87E-03  &  7.87E-03  &  7.87E-03  &  7.87E-03 &   7.87E-03      \\
 Na    &  5.10E-05  &  5.10E-05  &  5.10E-05  &  5.10E-05  &  5.10E-05  &  5.10E-05 &   5.10E-05      \\
 Mg    &  1.76E-02  &  1.76E-02  &  1.76E-02  &  1.76E-02  &  1.76E-02  &  1.76E-02 &   1.76E-02      \\
 Al    &  7.06E-05  &  7.06E-05  &  7.06E-05  &  7.06E-05  &  7.07E-05  &  7.07E-05 &   7.07E-05      \\
 Si    &  4.08E-02  &  5.44E-02  &  6.02E-02  &  6.16E-02  &  6.16E-02  &  6.16E-02 &   6.16E-02      \\
 P     &  1.65E-05  &  1.68E-05  &  1.70E-05  &  1.72E-05  &  1.73E-05  &  1.75E-05 &   1.75E-05      \\
 S     &  2.22E-02  &  3.10E-02  &  3.55E-02  &  3.69E-02  &  3.69E-02  &  3.69E-02 &   3.69E-02      \\
 Cl    &  1.59E-05  &  1.59E-05  &  1.59E-05  &  1.71E-05  &  1.97E-05  &  3.09E-05 &   3.16E-05      \\
 Ar    &  4.07E-03  &  5.73E-03  &  6.73E-03  &  7.19E-03  &  7.20E-03  &  7.20E-03 &   7.21E-03      \\
 K     &  7.65E-06  &  7.67E-06  &  7.68E-06  &  7.69E-06  &  7.70E-06  &  7.82E-06 &   7.83E-06      \\
 Ca    &  3.53E-03  &  5.00E-03  &  6.06E-03  &  6.81E-03  &  6.85E-03  &  6.93E-03 &   6.96E-03      \\
 Sc    &  3.85E-08  &  4.06E-08  &  4.21E-08  &  6.02E-07  &  1.90E-06  &  1.11E-05 &   1.15E-05      \\
 Ti    &  6.19E-06  &  2.50E-05  &  4.41E-05  &  8.95E-05  &  1.37E-04  &  2.19E-04 &   2.54E-04      \\
 V     &  2.16E-07  &  6.38E-07  &  8.96E-07  &  9.72E-07  &  9.72E-07  &  9.73E-07 &   9.73E-07      \\
 Cr    &  5.44E-05  &  3.13E-04  &  6.06E-04  &  1.14E-03  &  1.20E-03  &  1.32E-03 &   1.36E-03      \\
 Mn    &  1.35E-05  &  3.60E-05  &  5.14E-05  &  5.76E-05  &  5.76E-05  &  5.76E-05 &   5.76E-05      \\
 Fe    &  1.15E-03  &  5.25E-03  &  1.03E-02  &  5.08E-02  &  7.63E-02  &  1.02E-01 &   1.18E-01      \\
 Co    &  4.12E-08  &  4.25E-08  &  4.31E-08  &  1.32E-04  &  3.42E-04  &  6.40E-04 &   1.15E-03      \\
 Ni    &  2.77E-05  &  4.93E-05  &  6.60E-05  &  6.09E-04  &  1.18E-03  &  2.17E-03 &   2.43E-03      \\
\hline
\end{tabular}

\end{table*}

\begin{table*}
\caption{\bf Yields of the $\rm \bf 20~M_\odot$ Z=0 model.}
\begin{tabular}{lccccccc}
\hline
 $\rm M(^{56}Ni)_{}^{}$ &    0.0010  &    0.0100  &    0.0500  &    0.1000 &     0.1500  &    0.2000&      0.2226    \\
\hline
 $\rm M_{\rm cut}$  &      2.12  &      2.04  &      1.96  &      1.90 &       1.84  &      1.75&        1.72    \\
 $\rm M_{\rm ejected}$   &     17.88  &     17.96  &     18.04  &     18.10 &      18.16  &     18.25&       18.28    \\
 H     &  9.75E+00  &  9.75E+00  &  9.75E+00  &  9.75E+00 &   9.75E+00  &  9.75E+00&    9.75E+00    \\
 He    &  6.22E+00  &  6.22E+00  &  6.22E+00  &  6.22E+00 &   6.22E+00  &  6.22E+00&    6.22E+00    \\
 C     &  2.95E-01  &  2.95E-01  &  2.95E-01  &  2.95E-01 &   2.95E-01  &  2.95E-01&    2.95E-01    \\
 N     &  2.25E-05  &  2.25E-05  &  2.25E-05  &  2.25E-05 &   2.25E-05  &  2.25E-05&    2.25E-05    \\
 O     &  1.07E+00  &  1.07E+00  &  1.07E+00  &  1.07E+00 &   1.07E+00  &  1.07E+00&    1.07E+00    \\
 F     &  1.06E-08  &  1.06E-08  &  1.06E-08  &  1.06E-08 &   1.06E-08  &  1.06E-08&    1.06E-08    \\
 Ne    &  2.62E-02  &  2.62E-02  &  2.62E-02  &  2.62E-02 &   2.62E-02  &  2.62E-02&    2.62E-02    \\
 Na    &  1.53E-04  &  1.53E-04  &  1.53E-04  &  1.53E-04 &   1.53E-04  &  1.53E-04&    1.53E-04    \\
 Mg    &  4.70E-02  &  4.70E-02  &  4.70E-02  &  4.70E-02 &   4.70E-02  &  4.70E-02&    4.70E-02    \\
 Al    &  2.56E-04  &  2.56E-04  &  2.56E-04  &  2.56E-04 &   2.57E-04  &  2.57E-04&    2.57E-04    \\
 Si    &  9.13E-02  &  1.30E-01  &  1.46E-01  &  1.46E-01 &   1.46E-01  &  1.46E-01&    1.46E-01    \\
 P     &  7.77E-05  &  7.83E-05  &  7.90E-05  &  7.90E-05 &   7.91E-05  &  7.93E-05&    7.93E-05    \\
 S     &  4.74E-02  &  7.30E-02  &  8.65E-02  &  8.65E-02 &   8.65E-02  &  8.65E-02&    8.65E-02    \\
 Cl    &  4.68E-05  &  4.70E-05  &  4.70E-05  &  4.73E-05 &   4.87E-05  &  6.50E-05&    6.85E-05    \\
 Ar    &  8.30E-03  &  1.32E-02  &  1.66E-02  &  1.66E-02 &   1.66E-02  &  1.66E-02&    1.66E-02    \\
 K     &  1.97E-05  &  1.97E-05  &  1.98E-05  &  1.98E-05 &   1.98E-05  &  2.00E-05&    2.00E-05    \\
 Ca    &  6.99E-03  &  1.13E-02  &  1.53E-02  &  1.54E-02 &   1.54E-02  &  1.55E-02&    1.56E-02    \\
 Sc    &  8.84E-08  &  9.40E-08  &  9.98E-08  &  2.20E-07 &   8.64E-07  &  1.59E-05&    1.84E-05    \\
 Ti    &  8.93E-06  &  5.55E-05  &  1.47E-04  &  1.77E-04 &   2.45E-04  &  3.68E-04&    4.05E-04    \\
 V     &  3.81E-07  &  1.67E-06  &  3.10E-06  &  3.10E-06 &   3.10E-06  &  3.10E-06&    3.10E-06    \\
 Cr    &  6.15E-05  &  6.72E-04  &  2.42E-03  &  2.46E-03 &   2.56E-03  &  2.73E-03&    2.78E-03    \\
 Mn    &  2.02E-05  &  8.34E-05  &  1.83E-04  &  1.83E-04 &   1.83E-04  &  1.83E-04&    1.83E-04    \\
 Fe    &  1.30E-03  &  1.06E-02  &  5.10E-02  &  1.02E-01 &   1.53E-01  &  2.04E-01&    2.27E-01    \\
 Co    &  8.64E-08  &  8.89E-08  &  9.11E-08  &  1.81E-04 &   5.21E-04  &  8.44E-04&    9.98E-04    \\
 Ni    &  3.95E-05  &  8.03E-05  &  1.52E-04  &  6.88E-04 &   1.43E-03  &  3.22E-03&    3.94E-03    \\
\hline
\end{tabular}

\end{table*}

\begin{table*}
\caption{\bf Yields of the $\rm \bf 25~M_\odot$ Z=0 model.}
\begin{tabular}{lccccccc}
\hline
 $\rm M(^{56}Ni)_{}^{}$  &    0.0010  &    0.0100  &    0.0500  &    0.1000 &     0.1500  &    0.2000   &   0.3288   \\
\hline
 $\rm M_{\rm cut}$  &      2.50  &      2.40  &      2.30  &      2.24 &       2.18  &      2.12   &     1.93   \\
 $\rm M_{\rm ejected}$   &     22.50  &     22.60  &     22.70  &     22.76 &      22.82  &     22.88   &    23.07   \\
 H     &  1.14E+01  &  1.14E+01  &  1.14E+01  &  1.14E+01 &   1.14E+01  &  1.14E+01   & 1.14E+01   \\
 He    &  7.85E+00  &  7.85E+00  &  7.85E+00  &  7.85E+00 &   7.85E+00  &  7.85E+00   & 7.85E+00   \\
 C     &  4.13E-01  &  4.13E-01  &  4.13E-01  &  4.13E-01 &   4.13E-01  &  4.13E-01   & 4.13E-01   \\
 N     &  6.54E-02  &  6.54E-02  &  6.54E-02  &  6.54E-02 &   6.54E-02  &  6.54E-02   & 6.54E-02   \\
 O     &  2.05E+00  &  2.05E+00  &  2.05E+00  &  2.05E+00 &   2.05E+00  &  2.05E+00   & 2.05E+00   \\
 F     &  8.73E-08  &  8.73E-08  &  8.73E-08  &  8.73E-08 &   8.73E-08  &  8.73E-08   & 8.73E-08   \\
 Ne    &  4.27E-02  &  4.27E-02  &  4.27E-02  &  4.27E-02 &   4.27E-02  &  4.27E-02   & 4.27E-02   \\
 Na    &  2.34E-04  &  2.34E-04  &  2.34E-04  &  2.34E-04 &   2.34E-04  &  2.34E-04   & 2.34E-04   \\
 Mg    &  6.78E-02  &  6.78E-02  &  6.78E-02  &  6.78E-02 &   6.78E-02  &  6.78E-02   & 6.78E-02   \\
 Al    &  3.39E-04  &  3.39E-04  &  3.39E-04  &  3.39E-04 &   3.39E-04  &  3.39E-04   & 3.39E-04   \\
 Si    &  1.23E-01  &  1.69E-01  &  1.96E-01  &  1.96E-01 &   1.96E-01  &  1.96E-01   & 1.96E-01   \\
 P     &  7.07E-05  &  7.15E-05  &  7.29E-05  &  7.29E-05 &   7.30E-05  &  7.31E-05   & 7.34E-05   \\
 S     &  6.44E-02  &  9.51E-02  &  1.17E-01  &  1.17E-01 &   1.17E-01  &  1.17E-01   & 1.17E-01   \\
 Cl    &  4.79E-05  &  4.80E-05  &  4.81E-05  &  4.84E-05 &   4.98E-05  &  5.13E-05   & 7.75E-05   \\
 Ar    &  1.14E-02  &  1.74E-02  &  2.24E-02  &  2.27E-02 &   2.27E-02  &  2.27E-02   & 2.27E-02   \\
 K     &  1.95E-05  &  1.95E-05  &  1.96E-05  &  1.96E-05 &   1.96E-05  &  1.96E-05   & 1.99E-05   \\
 Ca    &  9.82E-03  &  1.51E-02  &  2.07E-02  &  2.12E-02 &   2.12E-02  &  2.13E-02   & 2.15E-02   \\
 Sc    &  1.23E-07  &  1.30E-07  &  1.37E-07  &  2.55E-07 &   9.48E-07  &  1.63E-06   & 2.75E-05   \\
 Ti    &  1.07E-05  &  5.99E-05  &  1.77E-04  &  2.05E-04 &   2.50E-04  &  3.06E-04   & 5.23E-04   \\
 V     &  3.08E-07  &  1.41E-06  &  2.66E-06  &  2.72E-06 &   2.72E-06  &  2.72E-06   & 2.72E-06   \\
 Cr    &  5.89E-05  &  6.73E-04  &  2.73E-03  &  3.31E-03 &   3.37E-03  &  3.45E-03   & 3.75E-03   \\
 Mn    &  1.78E-05  &  7.24E-05  &  1.48E-04  &  1.55E-04 &   1.55E-04  &  1.55E-04   & 1.55E-04   \\
 Fe    &  1.24E-03  &  1.05E-02  &  5.08E-02  &  1.01E-01 &   1.52E-01  &  2.03E-01   & 3.33E-01   \\
 Co    &  1.20E-07  &  1.22E-07  &  1.24E-07  &  4.18E-05 &   1.39E-04  &  3.33E-04   & 8.31E-04   \\
 Ni    &  4.37E-05  &  8.22E-05  &  1.40E-04  &  6.21E-04 &   1.56E-03  &  2.41E-03   & 6.56E-03   \\
\hline
\end{tabular}

\end{table*}

\begin{table*}
\caption{\bf Yields of the $\rm \bf 35~M_\odot$ Z=0 model.}

\begin{tabular}{lccccccc}
\hline
 $\rm M(^{56}Ni)_{}^{}$  &    0.0010  &    0.0100  &    0.0500  &    0.1000 &     0.1500  &    0.2000   &   0.6422   \\
\hline
 $\rm M_{\rm cut}$  &      2.96  &      2.84  &      2.67  &      2.60 &       2.54  &      2.49   &     1.93   \\
 $\rm M_{\rm ejected}$   &     32.04  &     32.16  &     32.33  &     32.40 &      32.46  &     32.51   &    33.07   \\
 H     &  1.47E+01  &  1.47E+01  &  1.47E+01  &  1.47E+01 &   1.47E+01  &  1.47E+01   & 1.47E+01   \\
 He    &  1.06E+01  &  1.06E+01  &  1.06E+01  &  1.06E+01 &   1.06E+01  &  1.06E+01   & 1.06E+01   \\
 C     &  5.08E-01  &  5.08E-01  &  5.08E-01  &  5.08E-01 &   5.08E-01  &  5.08E-01   & 5.08E-01   \\
 N     &  3.82E-04  &  3.82E-04  &  3.82E-04  &  3.82E-04 &   3.82E-04  &  3.82E-04   & 3.82E-04   \\
 O     &  4.93E+00  &  4.93E+00  &  4.93E+00  &  4.93E+00 &   4.93E+00  &  4.93E+00   & 4.93E+00   \\
 F     &  3.63E-07  &  3.63E-07  &  3.63E-07  &  3.63E-07 &   3.63E-07  &  3.63E-07   & 3.63E-07   \\
 Ne    &  3.33E-01  &  3.33E-01  &  3.33E-01  &  3.33E-01 &   3.33E-01  &  3.33E-01   & 3.33E-01   \\
 Na    &  8.55E-04  &  8.55E-04  &  8.55E-04  &  8.55E-04 &   8.55E-04  &  8.55E-04   & 8.55E-04   \\
 Mg    &  1.02E-01  &  1.02E-01  &  1.02E-01  &  1.02E-01 &   1.02E-01  &  1.02E-01   & 1.02E-01   \\
 Al    &  5.94E-04  &  5.94E-04  &  5.94E-04  &  5.94E-04 &   5.94E-04  &  5.94E-04   & 5.94E-04   \\
 Si    &  1.41E-01  &  1.97E-01  &  2.59E-01  &  2.65E-01 &   2.66E-01  &  2.66E-01   & 2.66E-01   \\
 P     &  5.63E-05  &  5.74E-05  &  5.98E-05  &  6.05E-05 &   6.05E-05  &  6.06E-05   & 6.09E-05   \\
 S     &  7.78E-02  &  1.17E-01  &  1.61E-01  &  1.68E-01 &   1.68E-01  &  1.68E-01   & 1.68E-01   \\
 Cl    &  4.34E-05  &  4.36E-05  &  4.38E-05  &  4.38E-05 &   4.41E-05  &  4.46E-05   & 6.66E-05   \\
 Ar    &  1.43E-02  &  2.20E-02  &  3.14E-02  &  3.34E-02 &   3.35E-02  &  3.35E-02   & 3.36E-02   \\
 K     &  1.93E-05  &  1.94E-05  &  1.95E-05  &  1.95E-05 &   1.95E-05  &  1.95E-05   & 1.99E-05   \\
 Ca    &  1.28E-02  &  1.99E-02  &  2.92E-02  &  3.22E-02 &   3.24E-02  &  3.24E-02   & 3.28E-02   \\
 Sc    &  1.69E-07  &  1.78E-07  &  1.88E-07  &  1.92E-07 &   2.92E-07  &  5.37E-07   & 2.41E-05   \\
 Ti    &  1.19E-05  &  6.37E-05  &  2.20E-04  &  3.17E-04 &   3.33E-04  &  3.62E-04   & 7.99E-04   \\
 V     &  2.77E-07  &  1.26E-06  &  3.17E-06  &  3.83E-06 &   3.89E-06  &  3.89E-06   & 3.89E-06   \\
 Cr    &  5.65E-05  &  6.58E-04  &  3.07E-03  &  5.18E-03 &   5.41E-03  &  5.45E-03   & 6.05E-03   \\
 Mn    &  1.76E-05  &  6.81E-05  &  1.73E-04  &  2.28E-04 &   2.34E-04  &  2.34E-04   & 2.34E-04   \\
 Fe    &  1.23E-03  &  1.05E-02  &  5.09E-02  &  1.01E-01 &   1.52E-01  &  2.02E-01   & 6.56E-01   \\
 Co    &  2.15E-07  &  2.17E-07  &  2.19E-07  &  2.20E-07 &   5.72E-05  &  1.29E-04   & 3.22E-03   \\
 Ni    &  5.84E-05  &  9.84E-05  &  1.67E-04  &  2.09E-04 &   8.29E-04  &  1.59E-03   & 7.16E-03   \\
\hline
\end{tabular}

\end{table*}

\begin{table*}
\caption{\bf Yields of the $\rm \bf 50~M_\odot$ Z=0 model.}

\begin{tabular}{lccccccc}
\hline
 $\rm M(^{56}Ni)_{}^{}$  &    0.0010  &    0.0100  &    0.0500  &    0.1000 &     0.1500  &    0.2000 &   1.0198   \\
\hline
 $\rm M_{\rm cut}$  &      3.83  &      3.68  &      3.41  &      3.29 &       3.23  &      3.17 &     2.18   \\
 $\rm M_{\rm ejected}$   &     46.17  &     46.32  &     46.59  &     46.71 &      46.77  &     46.83 &    47.82   \\
 H     &  1.94E+01  &  1.94E+01  &  1.94E+01  &  1.94E+01 &   1.94E+01  &  1.94E+01 & 1.94E+01   \\
 He    &  1.51E+01  &  1.51E+01  &  1.51E+01  &  1.51E+01 &   1.51E+01  &  1.51E+01 & 1.51E+01   \\
 C     &  1.08E+00  &  1.08E+00  &  1.08E+00  &  1.08E+00 &   1.08E+00  &  1.08E+00 & 1.08E+00   \\
 N     &  8.14E-07  &  8.14E-07  &  8.14E-07  &  8.14E-07 &   8.14E-07  &  8.14E-07 & 8.14E-07   \\
 O     &  8.97E+00  &  8.97E+00  &  8.97E+00  &  8.97E+00 &   8.97E+00  &  8.97E+00 & 8.97E+00   \\
 F     &  2.47E-09  &  2.47E-09  &  2.47E-09  &  2.47E-09 &   2.47E-09  &  2.47E-09 & 2.47E-09   \\
 Ne    &  3.24E-01  &  3.24E-01  &  3.24E-01  &  3.24E-01 &   3.24E-01  &  3.24E-01 & 3.24E-01   \\
 Na    &  5.42E-04  &  5.42E-04  &  5.42E-04  &  5.42E-04 &   5.42E-04  &  5.42E-04 & 5.42E-04   \\
 Mg    &  1.31E-01  &  1.31E-01  &  1.31E-01  &  1.31E-01 &   1.31E-01  &  1.31E-01 & 1.31E-01   \\
 Al    &  5.41E-04  &  5.41E-04  &  5.41E-04  &  5.41E-04 &   5.41E-04  &  5.41E-04 & 5.41E-04   \\
 Si    &  2.08E-01  &  2.77E-01  &  3.99E-01  &  4.27E-01 &   4.30E-01  &  4.30E-01 & 4.30E-01   \\
 P     &  9.37E-05  &  9.49E-05  &  9.83E-05  &  9.99E-05 &   1.00E-04  &  1.00E-04 & 1.01E-04   \\
 S     &  1.18E-01  &  1.68E-01  &  2.51E-01  &  2.74E-01 &   2.78E-01  &  2.78E-01 & 2.78E-01   \\
 Cl    &  7.36E-05  &  7.39E-05  &  7.42E-05  &  7.43E-05 &   7.44E-05  &  7.44E-05 & 1.13E-04   \\
 Ar    &  2.17E-02  &  3.19E-02  &  4.76E-02  &  5.34E-02 &   5.46E-02  &  5.49E-02 & 5.50E-02   \\
 K     &  3.30E-05  &  3.31E-05  &  3.32E-05  &  3.33E-05 &   3.33E-05  &  3.33E-05 & 3.41E-05   \\
 Ca    &  1.92E-02  &  2.92E-02  &  4.34E-02  &  5.01E-02 &   5.22E-02  &  5.28E-02 & 5.34E-02   \\
 Sc    &  2.93E-07  &  3.08E-07  &  3.21E-07  &  3.28E-07 &   3.31E-07  &  3.33E-07 & 5.08E-05   \\
 Ti    &  1.40E-05  &  6.89E-05  &  2.63E-04  &  4.09E-04 &   4.88E-04  &  5.20E-04 & 1.20E-03   \\
 V     &  3.03E-07  &  1.27E-06  &  4.41E-06  &  6.13E-06 &   6.91E-06  &  7.30E-06 & 7.30E-06   \\
 Cr    &  5.55E-05  &  6.31E-04  &  3.32E-03  &  5.92E-03 &   7.85E-03  &  8.93E-03 & 9.88E-03   \\
 Mn    &  2.00E-05  &  7.20E-05  &  2.32E-04  &  3.43E-04 &   4.17E-04  &  4.70E-04 & 4.70E-04   \\
 Fe    &  1.30E-03  &  1.05E-02  &  5.12E-02  &  1.02E-01 &   1.52E-01  &  2.02E-01 & 1.04E+00   \\
 Co    &  4.48E-07  &  4.50E-07  &  4.54E-07  &  4.55E-07 &   4.56E-07  &  1.09E-05 & 3.90E-03   \\
 Ni    &  9.37E-05  &  1.39E-04  &  2.36E-04  &  2.93E-04 &   3.34E-04  &  4.42E-04 & 1.05E-02   \\
\hline
\end{tabular}

\end{table*}

\begin{table*}
\caption{\bf Yields of the $\rm \bf 80~M_\odot$ Z=0 model.}

\begin{tabular}{lccccccc}
\hline
 $\rm M(^{56}Ni)_{}^{}$  &    0.0010  &    0.0100  &    0.0500  &    0.1000 &     0.1500  &    0.2000 &   2.1325   \\
\hline
 $\rm M_{\rm cut}$  &      5.99  &      5.71  &      5.30  &      5.01 &       4.85  &      4.75 &     2.39   \\
 $\rm M_{\rm ejected}$  &     74.01  &     74.29  &     74.70  &     74.99 &      75.15  &     75.25 &    77.61   \\
 H     &  2.67E+01  &  2.67E+01  &  2.67E+01  &  2.67E+01 &   2.67E+01  &  2.67E+01 & 2.67E+01   \\
 He    &  2.45E+01  &  2.45E+01  &  2.45E+01  &  2.45E+01 &   2.45E+01  &  2.45E+01 & 2.45E+01   \\
 C     &  1.41E+00  &  1.41E+00  &  1.41E+00  &  1.41E+00 &   1.41E+00  &  1.41E+00 & 1.41E+00   \\
 N     &  9.63E-07  &  9.63E-07  &  9.63E-07  &  9.63E-07 &   9.63E-07  &  9.63E-07 & 9.63E-07   \\
 O     &  1.80E+01  &  1.80E+01  &  1.80E+01  &  1.80E+01 &   1.80E+01  &  1.80E+01 & 1.80E+01   \\
 F     &  2.84E-09  &  2.84E-09  &  2.84E-09  &  2.84E-09 &   2.84E-09  &  2.84E-09 & 2.84E-09   \\
 Ne    &  1.08E+00  &  1.08E+00  &  1.08E+00  &  1.08E+00 &   1.08E+00  &  1.08E+00 & 1.08E+00   \\
 Na    &  8.31E-04  &  8.31E-04  &  8.31E-04  &  8.31E-04 &   8.31E-04  &  8.31E-04 & 8.31E-04   \\
 Mg    &  3.23E-01  &  3.23E-01  &  3.23E-01  &  3.23E-01 &   3.23E-01  &  3.23E-01 & 3.23E-01   \\
 Al    &  8.44E-04  &  8.44E-04  &  8.44E-04  &  8.44E-04 &   8.44E-04  &  8.44E-04 & 8.45E-04   \\
 Si    &  4.36E-01  &  5.65E-01  &  7.58E-01  &  8.77E-01 &   9.22E-01  &  9.43E-01 & 9.65E-01   \\
 P     &  2.96E-04  &  2.97E-04  &  3.00E-04  &  3.03E-04 &   3.04E-04  &  3.05E-04 & 3.08E-04   \\
 S     &  2.42E-01  &  3.44E-01  &  4.70E-01  &  5.54E-01 &   5.92E-01  &  6.11E-01 & 6.35E-01   \\
 Cl    &  1.56E-04  &  1.57E-04  &  1.58E-04  &  1.58E-04 &   1.58E-04  &  1.58E-04 & 2.11E-04   \\
 Ar    &  4.31E-02  &  6.54E-02  &  8.83E-02  &  1.05E-01 &   1.14E-01  &  1.19E-01 & 1.26E-01   \\
 K     &  7.51E-05  &  7.56E-05  &  7.58E-05  &  7.59E-05 &   7.60E-05  &  7.60E-05 & 7.90E-05   \\
 Ca    &  3.72E-02  &  6.09E-02  &  8.08E-02  &  9.66E-02 &   1.06E-01  &  1.12E-01 & 1.25E-01   \\
 Sc    &  6.75E-07  &  7.21E-07  &  7.40E-07  &  7.55E-07 &   7.65E-07  &  7.71E-07 & 8.68E-05   \\
 Ti    &  1.53E-05  &  8.79E-05  &  2.98E-04  &  5.24E-04 &   7.00E-04  &  8.39E-04 & 2.56E-03   \\
 V     &  4.29E-07  &  2.01E-06  &  6.87E-06  &  1.18E-05 &   1.51E-05  &  1.75E-05 & 2.45E-05   \\
 Cr    &  5.39E-05  &  5.90E-04  &  3.33E-03  &  6.59E-03 &   9.47E-03  &  1.20E-02 & 2.31E-02   \\
 Mn    &  3.12E-05  &  1.20E-04  &  3.59E-04  &  6.14E-04 &   8.07E-04  &  9.67E-04 & 1.59E-03   \\
 Fe    &  1.64E-03  &  1.12E-02  &  5.23E-02  &  1.03E-01 &   1.54E-01  &  2.05E-01 & 2.18E+00   \\
 Co    &  1.19E-06  &  1.20E-06  &  1.20E-06  &  1.21E-06 &   1.21E-06  &  1.21E-06 & 9.32E-03   \\
 Ni    &  2.10E-04  &  2.97E-04  &  4.33E-04  &  5.57E-04 &   6.40E-04  &  7.06E-04 & 2.85E-02   \\
\hline
\end{tabular}

\end{table*}

\end{document}